\documentclass[11pt]{article}
\usepackage{blois}

\def\Journal#1#2#3#4{{#1} {\bf #2}, #3 (#4)}

\def\NIMA{{\em Nucl. Instrum. Methods} A}

\def\PLB{{\em Phys. Lett.}  B}
\def\PRL{\em Phys. Rev. Lett.}
\def\PRD{{\em Phys. Rev.} D}

\def\JHEP{\em J. High Energy Phys.}

\def\be{\begin{equation}}
\def\ee{\end{equation}}
\def\bea{\begin{eqnarray}}
\def\eea{\end{eqnarray}}

\def\fb{~fb$^{-1}$~}

\def\pythia{{\textsc{pythia}}}
\def\sherpa{{\textsc{sherpa}}}

\newcommand{\Photo}{}

\begin{document}
\vspace*{4cm}
\title{SELECTED SPRING 2013 HEAVY FLAVOR, QCD, AND ELECTROWEAK PHYSICS RESULTS FROM THE TEVATRON}

\author{ J.L.~HOLZBAUER }
 \author{ For the D0 and CDF Collaborations }
\address{University of Mississippi, University, Mississippi 38677, USA}

\maketitle\abstracts{
With the full Tevatron data set collected and being analyzed, many new results have been recently released.  This includes heavy flavor physics studies such as CP violation parameter measurements with $B^{\pm} \rightarrow J/\psi K^{\pm}$ and $B^{\pm} \rightarrow J/\psi \pi^{\pm}$ and $D^0-\overline{D}^0$ mixing.  Of the QCD and electroweak results, photon plus heavy flavor measurements and a search for anomalous quartic gauge couplings will be reviewed.  These various studies help to clarify the agreement between data and physics models and to search for new physics.}

\section{Overview of the Tevatron and Experiments}
All the studies in this document use data collected from 1.96 TeV $p\overline{p}$ collisions produced by the Fermilab Tevatron Collider.  Operation of this machine ended in September of 2011 after providing about 10\fb of data to the D0 and CDF detectors.  Analyses in this document use between 8.7 and 10.4\fb of data, depending on the specific analysis requirements.  

D0~\cite{d0:overview} and CDF~\cite{cdf:overview} are multi-purpose detectors with inner tracker, calorimeter and muon systems, and are described in more detail elsewhere.  The D0 CP violation analysis benefits from regular reversals of magnet polarity and the symmetric nature of the D0 detector.  More generally, because the colliding particles are $p$ and $\overline{p}$, the initial states are CP symmetric, which benefits CP violation studies.  Also, a different $\sqrt{s}$ of 1.96 TeV means studies like photon plus heavy flavor can complement those done with data from other machines.

\section{CP Violation Parameters in $B^{\pm} \rightarrow J/\psi K^{\pm}$ and $B^{\pm} \rightarrow J/\psi \pi^{\pm}$}
A measurement of CP violation parameters in $B^{\pm} \rightarrow J/\psi K^{\pm}$ and $B^{\pm} \rightarrow J/\psi \pi^{\pm}$ has been performed by the D0 Collaboration~\cite{Abazov:2013cp}.  This analysis is particularly interesting because it is a clean test of CP violation.  We expect at most about 0.3\% asymmetry for $B^{\pm} \rightarrow J/\psi K^{\pm}$ from penguin loops and a few percent for $B^{\pm} \rightarrow J/\psi \pi^{\pm}$ in the standard model~\cite{Dun:1994,Hou:2006du,Hou:1999xv}. 

The raw asymmetry in each state is measured first and then corrected for the reconstruction asymmetry of $K^+$ and $K^-$ in the detector.  The raw asymmetry is defined, with a similar definition for $\pi$ events, as the difference in the number of $K^+$ and $K^-$ events over the sum of all events with a $K^{\pm}$.  A correction for the kaon asymmetry is then applied because while the $K^-$ can interact with detector material to form hyperons, there is no equivalent interaction for $K^+$.  Thus, the $K^+$ typically travels farther through the detector, leading to a higher reconstruction efficiency.  In most analyses, there are also pion or tracking asymmetry corrections.  However, because of the symmetry in the D0 detector and the regular reversals of magnet polarity, there is no impact from such asymmetries on the measurement.  This is not assumed, but instead demonstrated using independent channels.  More details are given in the analysis paper~\cite{Abazov:2013cp}.

We use a maximum likelihood fit to extract the raw $J/\psi K^{\pm}$ and $J/\psi \pi^{\pm}$ asymmetries, and a fit to the invariant mass M of $K^{\pm}\pi^{\mp}$ in $K^{*0}\rightarrow K^+\pi^-(K^-\pi^+)$ to extract the kaon asymmetry.  Dominant uncertainties include the statistical uncertainty (by far the most dominant) and the uncertainty on the kaon asymmetry estimate.  The final results are as follows:
$A^{J/\psi K} = [0.59\pm 0.36$(stat)$\pm0.07$(syst)$]\%$
and
$A^{J/ \psi \pi} = [-4.2 \pm 4.4$(stat)$\pm0.9$(syst)$]\%$.  
Both results are consistent with the standard model prediction and the $J/\psi K^{\pm}$ asymmetry measurement is the most precise to date.

\section{$D^0-\overline{D}^0$ Mixing}
The CDF Collaboration has observed $D^0-\overline{D}^0$ mixing with 6.1 Gaussian sigma significance~\cite{Aaltonen:2012d0d0}.  This is a confirmation of the recent LHCb mixing observation~\cite{Aaij:2012nva}.  The $D^0$ meson is the final neutral meson where mixing has been observed, and interestingly, it could potentially have a large new physics contribution~\cite{Golowich:2007ka}.  However, determining this will require additional inputs, since $D^0$ mixing is believed to have large contributions from hard-to-estimate long range processes that appear to dominate over calculable short-range contributions~\cite{Falk:2004wg}.

Mixing is determined from the time dependence of the right sign (RS) decay $D^0 \rightarrow K \pi^+$ and the wrong sign (WS) decay $D^0 \rightarrow K \pi^-$, where the flavor of the $D^0$ is determined from the decay chain $D^* \rightarrow D^0 \pi^+$.  Specific decay chains implicitly include the charge conjugate.  In the case of no mixing, this ratio has no decay time dependence, while the presence of mixing is signaled by a quadratic dependence on decay time (the approximation for small mixing).

\begin{figure}
\begin{minipage}{0.55\linewidth}
\centerline{\includegraphics[width=0.9\linewidth]{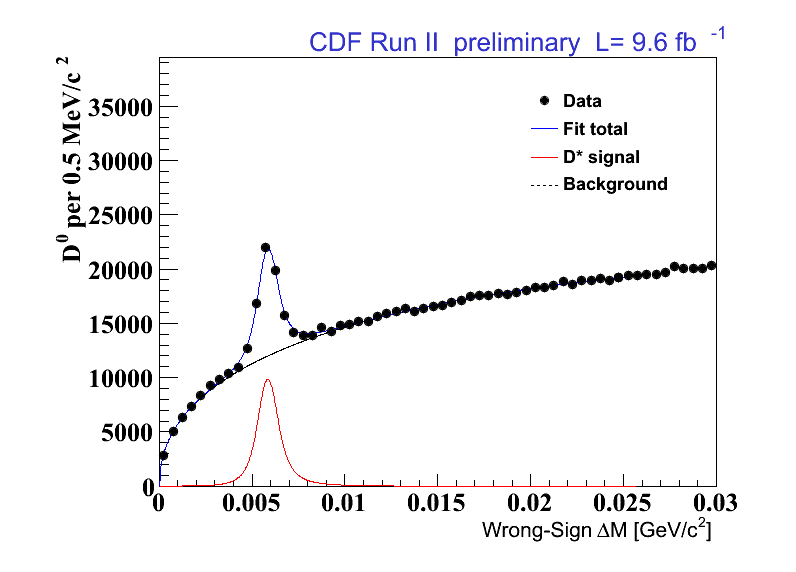}}
\end{minipage}
\hfill
\begin{minipage}{0.45\linewidth}
\centerline{\includegraphics[width=0.9\linewidth]{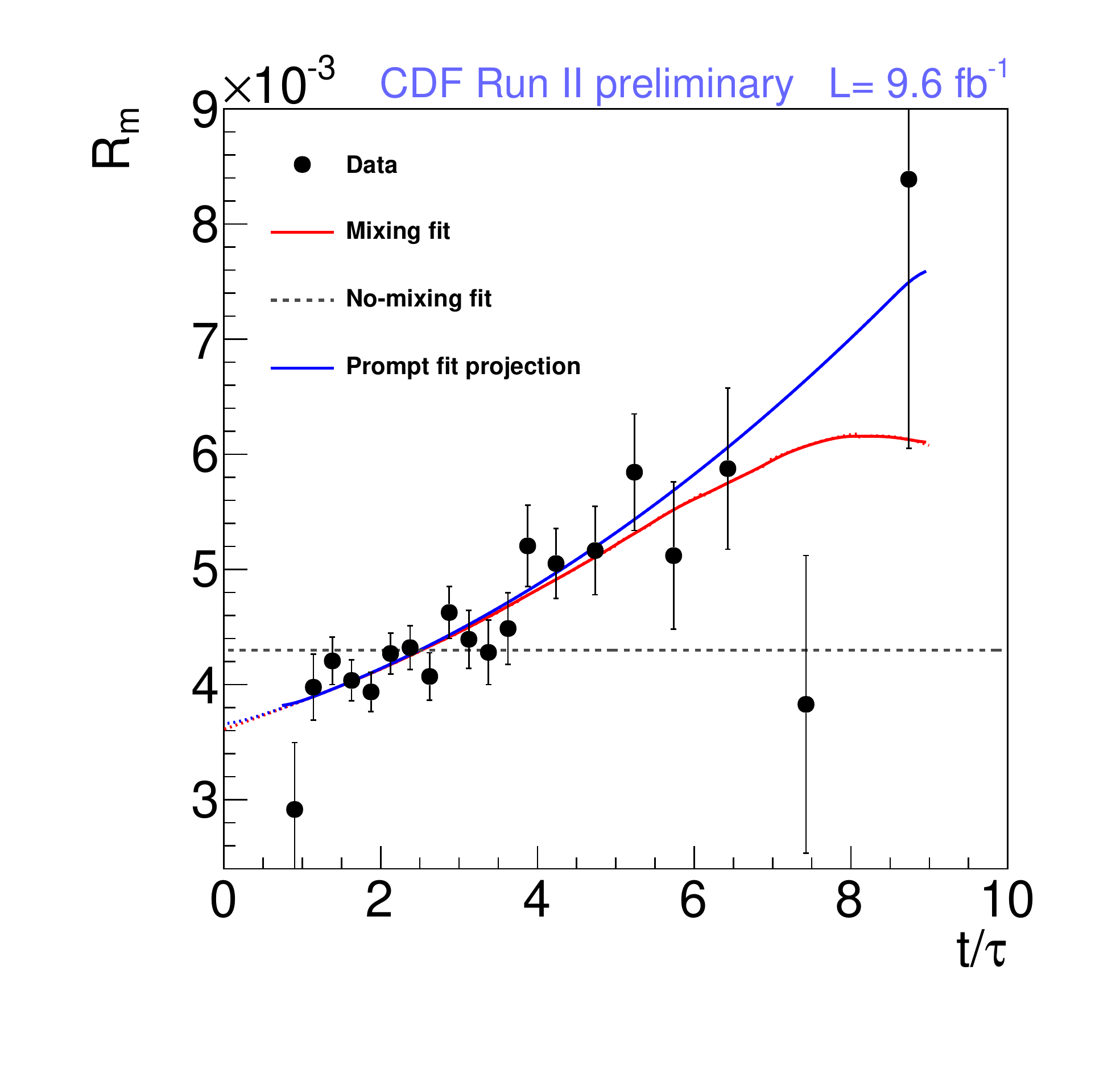}}
\end{minipage}
\caption{Distribution of WS $\Delta M$ (left) and ratio of WS to RS decays versus $t/\tau$ (right)}
\label{fig:d0d0}
\end{figure}

The same selection is applied to both RS and WS samples.  The selection is chosen to optimize the significance of the rarer WS decays using the scaled RS sample to estimate the WS signal, and sideband events to estimate backgrounds.  The RS and WS samples are divided into bins of $t/\tau$ and $\Delta M$, where $\tau$ is the mean $D^0$ lifetime and $\Delta M = M(K^+\pi^-\pi^+) - M(K^+\pi^-) - M(\pi^+)$.  The $D^0$ yield is determined in each bin from a fit to the $M(K\pi)$ distribution, and then the correctly tagged $D^0$'s from $D^*$'s are determined by a fit of the $\Delta M$ distribution (Figure~\ref{fig:d0d0}).  The ratio of WS to RS $D^0$ decays versus $t/\tau$ is shown in Figure~\ref{fig:d0d0}.  The time dependence can be clearly seen.  Bayesian and frequentist estimates show the no-mixing hypothesis is excluded at 6.1 Gaussian sigma.  Mixing parameters are available in the analysis note~\cite{Aaltonen:2012d0d0}.

\section{Photon plus Heavy Flavor}
Measurements of the photon plus heavy flavor processes have been made by both the D0~\cite{Abazov:2013cb} and CDF~\cite{Aaltonen:2012cb} Collaborations.  The D0 Collaboration has also performed a measurement of $\gamma+b$~\cite{Abazov:2012b}, which will not be discussed in detail here.  Photon plus heavy flavor processes are relatively clean processes which can be used to study the quark and gluon parton distribution functions (PDFs) and the rate of gluons splitting to quarks.  These processes are produced primarily through Compton-like scattering for photons with low $E_{T}$ (approximately $< 100~$GeV) and $q\overline{q}$ annihilation with gluon splitting otherwise.

In both sets of measurements, a central $\gamma$ is required, using a neural network (NN) to help identify it.  NN's are also used in a template fit to determine the rate of jets faking $\gamma$'s.  Additionally, the secondary vertex mass, the invariant mass of the charged particles near a secondary displaced vertex, is used in a template fit to determine the $b$ and $c$ quark fractions.

Distributions are produced and compared to a variety of event generators, including NLO~\cite{nloD0,nloCDF}, $k_{T}$ factorization~\cite{kt}, \sherpa~\cite{sherpa}, and \pythia~\cite{pythia}.  D0 also compared to a BHPS IC model~\cite{BHPS1,BHPS2} and a sea-like IC model~\cite{IC}.  In the BHPS IC model, the intrinsic charm quark generally has a high momentum fraction x, while in the sea-like IC model the charm PDF is similar to light-flavor sea quarks.  Each experiment used slightly different versions of most generators and PDF sets.  Please see the analysis notes for additional details~\cite{Abazov:2013cb,Aaltonen:2012cb}.  

\begin{figure}
\begin{minipage}{0.33\linewidth}
\centerline{\includegraphics[width=0.91\linewidth]{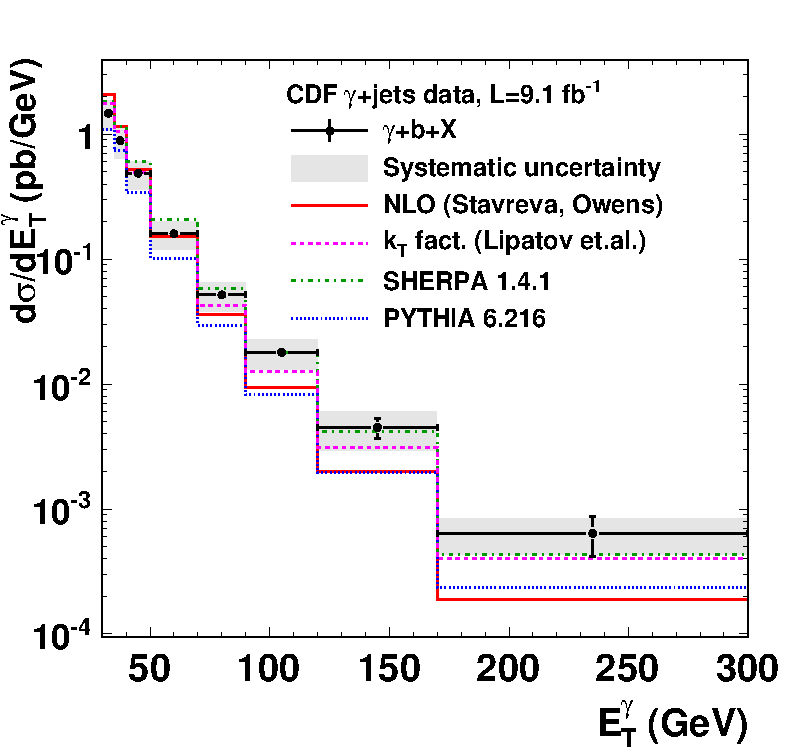}}
\end{minipage}
\hfill
\begin{minipage}{0.32\linewidth}
\centerline{\includegraphics[width=0.92\linewidth]{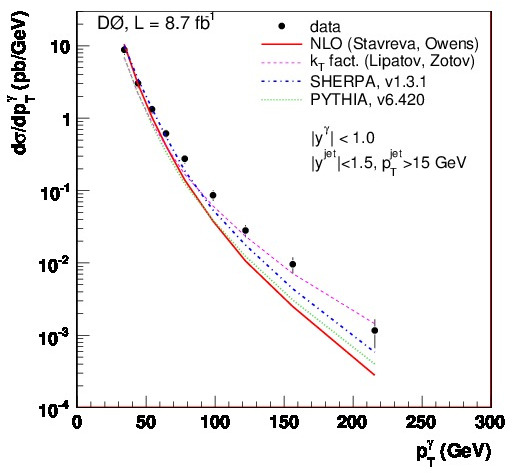}}
\end{minipage}
\hfill
\begin{minipage}{0.32\linewidth}
\centerline{\includegraphics[width=0.89\linewidth]{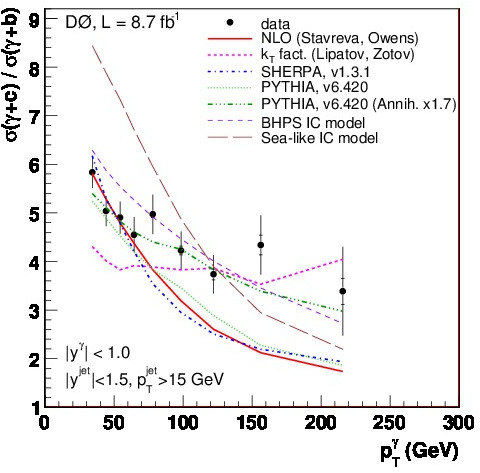}}
\end{minipage}
\caption{Distributions for $\gamma+b$ (left) $\gamma+c$ (center) and the ratio $\gamma+c/\gamma+b$ (right) }
\label{fig:hfcdf}
\end{figure}

Figure~\ref{fig:hfcdf} shows example distributions for the $\gamma+b$ process from CDF, the $\gamma+c$ process from D0, and the ratio $\gamma+c/\gamma+b$ from D0.  In the $\gamma+b$ distribution, we see that the \sherpa~or $k_{T}$ factorization agrees best with the data distribution for higher $E_{T}$ values, while the NLO in particular disagrees in this region.  In the $\gamma+c$ distribution we see better data and model agreement with $k_{T}$ factorization and somewhat with \sherpa~in the large $p_{T}$ region than with NLO, though $k_{T}$ factorization underestimates for low $p_{T}$.  Similarly, CDF sees disagreement with NLO for large $E_{T}$ and reasonable agreement with $k_{T}$ factorization and \sherpa.  In the distribution of the ratio $\gamma+c/\gamma+b$ from D0, we see improved data and model agreement for \pythia~if it is generated with a 1.7 enhancement of the rate of the annihilation process with gluon splitting to charm quarks.  CDF also sees better shape agreement in the $\gamma+c$ and $\gamma+b$ distributions if the gluon splitting to heavy flavor in \pythia~is enhanced by a factor of two.

\section{Anomalous Quartic Gauge Coupling Search}
The D0 Collaboration has performed a study looking for anomalous quartic gauge couplings (aQGC) $WW\gamma\gamma$ in events with an electron, positron, and missing transverse energy~\cite{Abazov:2013opa}.  This study is an extension of a Higgs boson search~\cite{Abazov:2013wha}.  An effective field theory~\cite{Belanger:aqgc} with dimension 6 operators is considered for the aQGC search, where the coupling constants under study are $a_0^W$ and $a_C^W$.  Both couplings are 0 in the standard model.  These couplings have not, until recently, been constrained beyond limits made by the OPAL Collaboration~\cite{Abbiendi:opal}.  This measurement was the first to do so.  Other recent results by the CMS collaboration improve the limits further~\cite{cms:aqgc}.

\begin{figure}
\begin{minipage}{0.32\linewidth}
\centerline{\includegraphics[width=0.9\linewidth]{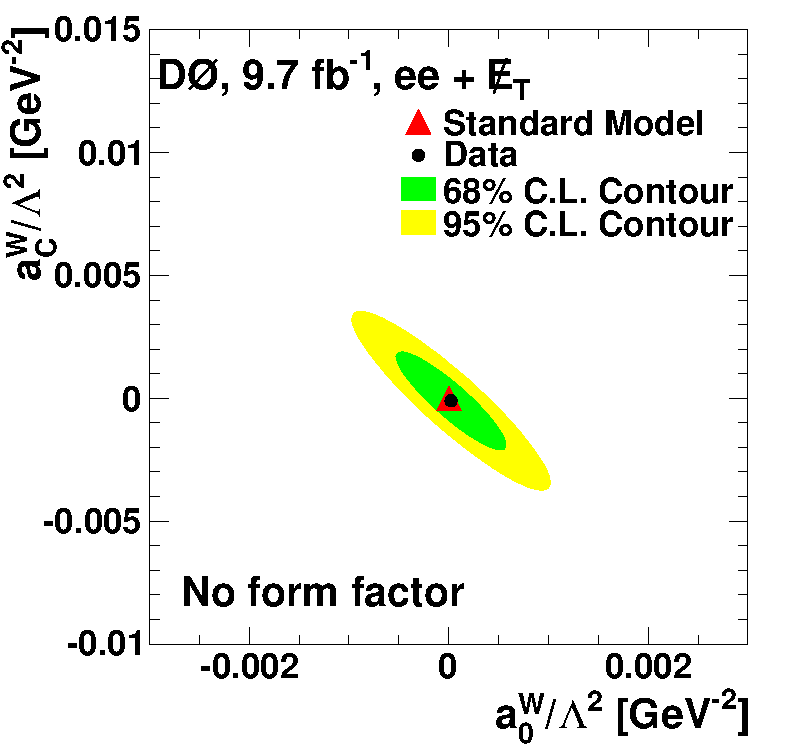}}
\end{minipage}
\hfill
\begin{minipage}{0.32\linewidth}
\centerline{\includegraphics[width=0.9\linewidth]{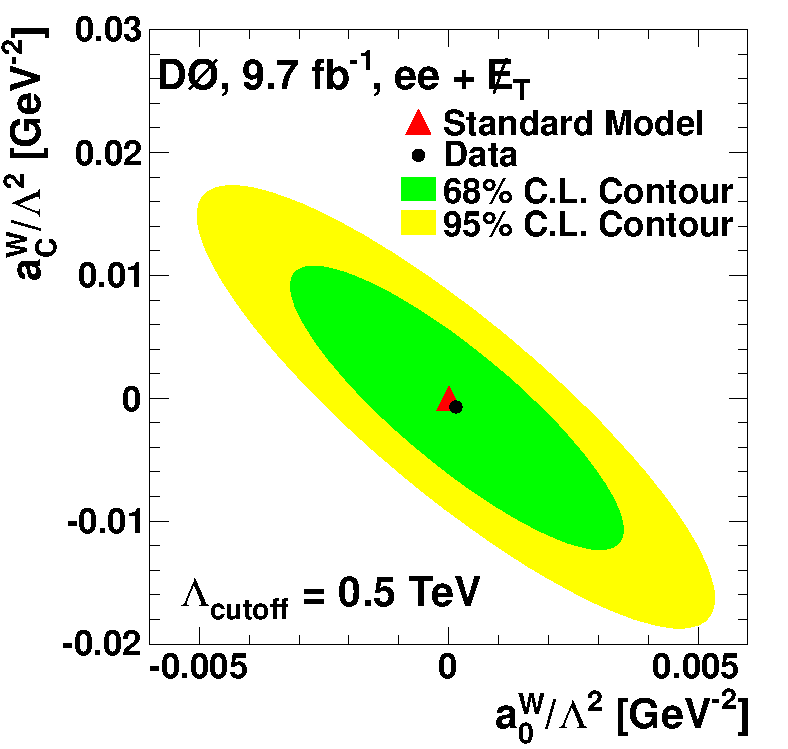}}
\end{minipage}
\hfill
\begin{minipage}{0.32\linewidth}
\centerline{\includegraphics[width=0.9\linewidth]{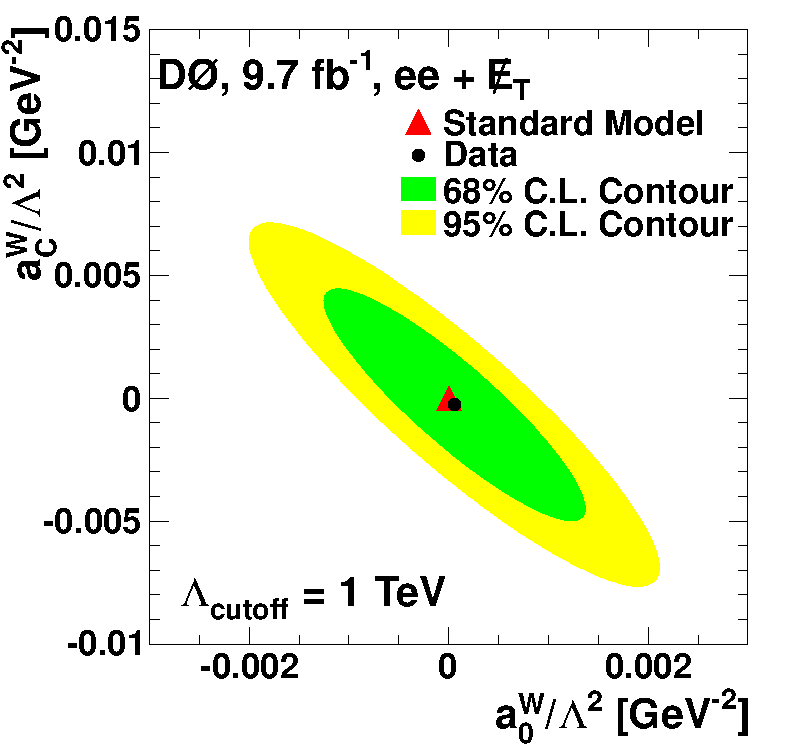}}
\end{minipage}
\caption{Limits on $a_0^W$ and $a_C^W$ for three different form factor values.  Left is without a form factor, middle is a form factor of 0.5~TeV, right is a form factor of 1.0~TeV.}
\label{fig:aqgc}
\end{figure}

The analysis is done as in the same way as the prior Higgs boson search but includes an extra jet veto.  The boosted decision tree (BDT) is trained for the aQGC signal in this analysis and the same BDT is used for both the $a_0^W$ and $a_C^W$ searches.  No significant excess above the background expectation is found and limits are determined for $\Lambda_{cutoff}$ of 1.0~TeV, 0.5~TeV and no form factor. The results are shown in Figure~\ref{fig:aqgc} and in particular the upper limits for $\Lambda_{cutoff}$ of 0.5~TeV at 95\% C.L. are $|a_0^W/\Lambda^2|<0.0025~$GeV$^{-2}$ and $|a_C^W/\Lambda^2|<0.0092~$GeV$^{-2}$.  These results are a factor of 4 to 8 better than the OPAL Collaboration limits.

\end{document}